\newcommand{\draft}{T}

\documentclass[12pt,twoside,draft]{article}

\usepackage{epsf} 	

\newcommand{\briefrefs}{T}

\if T\briefrefs

	\newcommand{\Journal}{J.}
	\newcommand{\JournalP}{J.}
	\newcommand{\Eng}{Eng.}
	
	\newcommand{\Soc}{Soc.}

\else

	\newcommand{\Journal}{Journal of the}
	\newcommand{\JournalP}{Journal}
	\newcommand{\Eng}{Engineering}
	
	\newcommand{\Soc}{Society}

\fi

\newcommand{\beqn}{\begin{equation}}
\newcommand{\eeqn}{\end{equation}}

\newcommand{\be}{\begin{equation}}
\newcommand{\ee}{\end{equation}}
\newcommand{\beqa}{\begin{eqnarray}}
\newcommand{\eeqa}{\end{eqnarray}}
\newenvironment{vectra}[1]%
	{\left[ 		
	\begin{array}{#1}}
	{\end{array} \right]}

\newcommand{\bv}{\begin{vectra}}
\newcommand{\ev}{\end{vectra}}

\newcommand{\ba}{\begin{array}}
\newcommand{\ea}{\end{array}}

\newcommand{\Comment}[1]{}

\newcommand{\figureLongCapTwoCol}[3]{
	\if F\draft
	\begin{figure}[htbp]
	\centerline{\psfig{figure=#2,width=3.25in}} 
	\caption{\protect {\small #3}}
	\label{#1} 
	\end{figure}
	\fi
}

\newcommand{\bit}{\begin{itemize}}
\newcommand{\eit}{\end{itemize}}
\newcommand{\bsl}{\begin{slide}{}}
\newcommand{\esl}{\end{slide}}

\title{Generalization of a 3-D resonator model \\ 
	for the simulation of spherical enclosures\footnote{Accepted for publication in Applied Signal Processing, 2000}}

\author{Davide Rocchesso \\
 {\small Universit\`a di Verona} \\
	{\small Dipartimento Scientifico e Tecnologico} \\
	{\small Strada Le Grazie 15,  I-37134  Verona}\\
	{\small {\tt rocchesso@sci.univr.it} } \\
 Pierre Dutilleux \\
	{\small ZKM~$|$~Zentrum f\"{u}r Kunst und Medientechnologie} \\
	{\small {Institute for Music and Acoustics}} \\
	{\small Lorenzstrasse 19, D-76135  Karlsruhe} \\
	{\small{\tt dutilleux@zkm.de} }
}

\begin{document}
\pagestyle{myheadings}
\markboth{Applied Signal Processing, 2000 - Springer V. - Accepted for Publication }{D. Rocchesso and P. Dutilleux: Generalization of a 3D-resonator model}

\if T\draft
	\addtolength{\baselineskip}{.5\baselineskip}
\fi

\bibliographystyle{ieee}

\maketitle


\begin{abstract}
{A rectangular enclosure has such an even distribution of resonances that it can be  
accurately and efficiently modelled using a feedback delay network. Conversely, a non rectangular shape such as a  
sphere has a distribution of resonances that challenges the construction of an efficient model.
        This work proposes an extension of the already known feedback delay network  
structure to model the resonant properties of a sphere.
        A specific frequency distribution of resonances can be approximated, up to a  
certain frequency, by inserting an allpass filter of moderate order after each delay line  
of a feedback delay network.
        The structure used for rectangular boxes is therefore augmented  
with a set of allpass filters allowing parametric control over the enclosure size and the boundary properties. 
	\Comment{The filter parameters can be directly associated with a  
control of ``roundness'' along the three axes, thus allowing to morph a box into a  
cylinder into a sphere.}
        This work was motivated by informal listening tests which have shown that it is  
possible to identify a basic shape just from the distribution of its audible  
resonances.
}
\end{abstract}

\section{Introduction}
\begin{figure}[ht]
\if T\draft
	\epsfxsize=9.0cm
\else
	\epsfxsize=7.0cm
\fi
	\centerline{\epsfbox{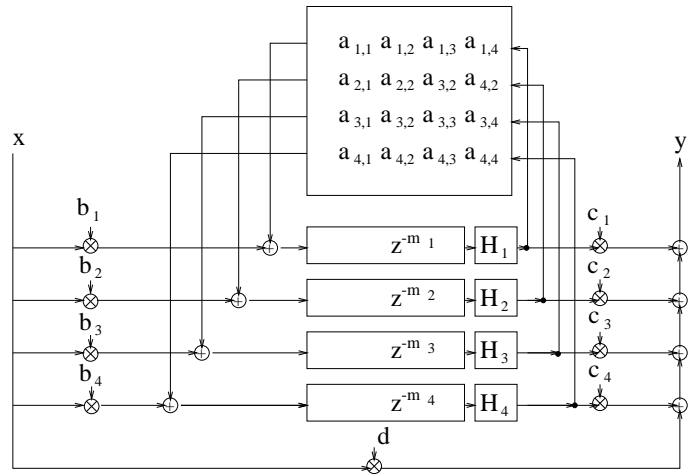}}
	\caption{ Feedback delay network of order 4.}
	\label{fdn}
\end{figure}

The feedback delay network (FDN) of order $N$, as depicted in fig.~\ref{fdn} for $N = 4$, is the multivariable generalization of the recursive comb filter, and it has been widely used to simulate the late reverbation of an enclosure~\cite{Jot92,jot,cmj95,rocjosieee}. The FDN with a diagonal matrix can be used to simulate a box with perfectly-reflecting walls. Energy absorption at the walls and in air can be accounted for by cascading the delay lines with  lowpass filters. In regular rooms, some energy gets transferred from one mode to another due to non-mirror reflection at the walls. This effect is called diffusion~\cite{Kuttruff} and is encompassed by the non-diagonal elements of the feedback matrix. As observed  in the time domain, diffusion produces a gradual increase in density of the impulse response. 
The delay lengths of a FDN are sometimes chosen with number-theoretic considerations in order to minimize the overlapping of echoes, as it was done with classic reverberation structures~\cite{MoorerReverb79}. A more physical criterion is based on the following observation. In a rectangular enclosure, the distribution of normal modes can be obtained as the composition of (infinite) harmonic series, each series being associated with the spatial direction of propagation of the plane wave fronts supporting the modes. For instance, the longitudinal size of a rectangular box is associated with a low-pitch mode and with all its multiples. Since any harmonic series of resonances can be reproduced by means of a recursive comb filter, a reference FDN can be constructed as a parallel connection of comb filters or, in other words, with a diagonal feedback matrix. For the rectangular enclosure, the delay lengths can be computed exactly from the geometry of the room~\cite{cmj95}. Specifically, given a limited number of delay units $N$, it is possible to determine the $N$ delay lengths such that the model reproduces all the resonances of the room up to a limit frequency that depends on $N$ and the dimensions of the room. Conversely, the number $N$ of delay units can be determined in such a way that the limit frequency coincides with the Schroeder frequency~\cite{SchroederLogan61}, above which the exact reproduction of resonances is not perceptually relevant and it is sufficient to keep the modal density constant.

The fact that different modes are differently excited according to the position of the sound source can be taken into account by a proper choice of the $\bf b$ coefficients in fig.~\ref{fdn}. Similarly, the pickup point should affect the choice of the $\bf c$ coefficients.

A model where normal modes, absorptions, and diffusion are all represented in the FDN structure has been called ``Ball within the Box'' (BaBo), since it can be visualized as a box of mirrors containing a single diffusing object (the Ball) represented by the matrix~\cite{cmj95}. Usually the model is completed with the explicit computation of early reflections that uses a geometric description of the acoustical scene.
\begin{figure*}[h]
\if T\draft
	\epsfxsize=12.0cm
\else
	\epsfxsize=10cm
\fi
	\centerline{\epsfbox{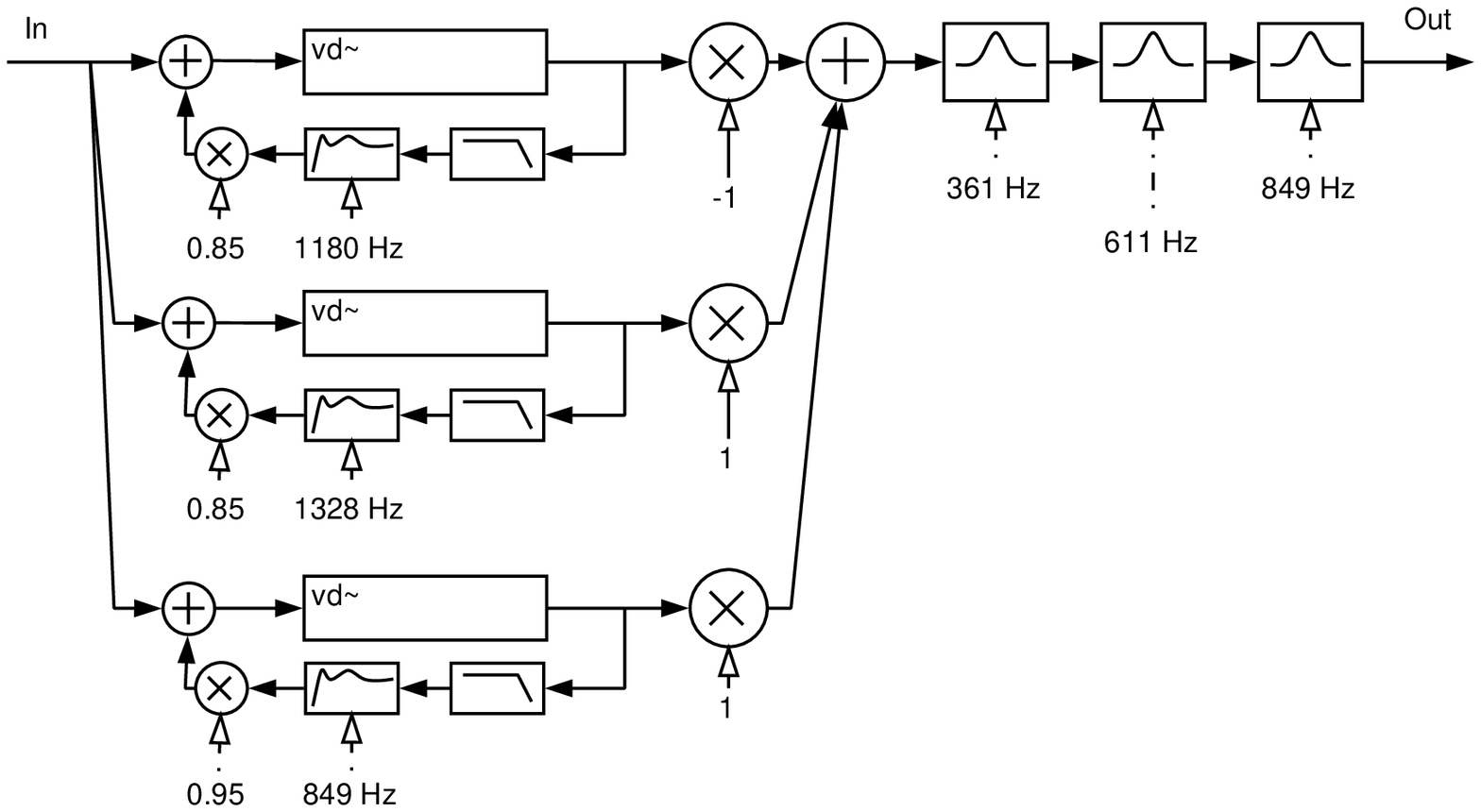}}
	\caption{A first attempt to imitate a spherical resonator of radius $a = 0.5 \mbox{m}$ at temperature
	$t = 13^\circ \mbox{C}$. Two filters are used in the feedback loop of each comb filter. 
	A 3rd order Chebychev highpass filter removes the
	lower resonances that do not match the roots of $j'_n(x) = 0$,
	and a first order lowpass filter accounts for  air absorption 
	and frequency-dependant reflections.
	}
	\label{BasicSphere}
\end{figure*}

Non-rectangular enclosures usually do not have an even distribution of resonances. In some relevant cases, however, the modal distribution can be calculated in closed form from the geometric specification of the enclosure. In particular, this paper deals with the spherical resonator, whose resonances can be found by computing the local extremal points of Bessel functions. The spherical Bessel functions tend to cosine functions for large values of the argument~\cite{MorseAndIngard}. 
A prior realization of the spherical resonator, depicted in fig.~\ref{BasicSphere}, exploited the fact that the extremal points are asymptotically equidistant, using recursive comb filters with feedback high-pass filters to reproduce the medium- and high-frequency resonances~\cite{Dutilleux99}. On the other hand, a set of  low-frequency resonances were individually  reproduced by tuned second-order resonant filters. Such prior realization was successfully experimented in the AML~$|$~Architecture and Music Laboratory, a museum installation where the visitor can experience how shapes such as, e.g. a tube, a cube or a sphere imprint a specific signature on the sounds. Informal reports from many listeners convinced us that it is indeed possible to identify basic shapes from the kind of resonance distribution they display. 

Whereas the models used within the AML are specific to each shape, we try here, starting from the  BaBo model, to design a single model valid for a few shapes. The BaBo model was initially designed for rectangular shapes but we extend it to the simulation of non-rectangular ones, in the hope that we can even feature a ``shape control handle''. This paper reports the extension of the BaBo model to spherical enclosures and the comparison of audible results with recordings made through acoustical resonators. We also indicate how the simulation of cylinders of various lengths and radii might be achieved by the same structure with the proper parameter design.

\section{Rectangular resonator model}
The BaBo model provides parametric control over the geometric and physical properties of a rectangular enclosure~\cite{cmj95}. The kernel of the model is a feedback delay network where the delay lines have length in seconds given by
\be
	d_{l,m,n} = \frac{2}{c \sqrt{(l/X)^2 + (m/Y)^2 + (n/Z)^2}}
\ee
where $c$ is the speed of sound and $l, m, n$ are triplets of small positive integers sharing no common (nontrivial) divisor. 
Triplets with two zeros correspond to axial modes. Triplets with one zero correspond to tangential modes. Triplets with no zero correspond to oblique modes~\cite{morse}.

If the feedback matrix is diagonal we have a parallel of comb filters and this corresponds to a perfectly reflecting enclosure. 
In this case each comb filter represents the triplets $(l,~m,~n)$, $(2l,~2m,~2n)$, $(3l,~3m,~3n)$, etc., thus giving a perfectly harmonic series of resonances.
In fact, each harmonic series of resonances is provided by a closed plane wave path propagating back and forth along a precise direction in space. If there is diffusion at the walls the plane wave fronts get scattered along many directions at every reflection. This diffusion phenomenon is reproduced in the BaBo model by  non-diagonal matrix coefficients. If the elements of the matrix have all the same magnitude we have perfect, so called Lambertian, diffusion. Since the comb filters feed each other as the wavefronts scatter along several directions, the harmonicity of each comb series is somewhat broken in a diffusive enclosure. 

Whereas the BaBo structure is well suited for imitating resonators characterized by overlapping series of harmonic resonances, it seems difficult to adapt it for the simulation of non-rectangular enclosures. Indeed, this is possible with moderate extra effort, as we will explain in section~\ref{sphmodel}.

\section{Acoustics of the sphere}
The modal frequencies $f_{ns}$ of a sphere are proportional to the roots $z_{ns}$ of 
\be
	j'_n (x)=0	\; , n = 0, 1, \dots					\label{equation}
\ee
 where $j_n$ is the spherical Bessel function\footnote{The spherical Bessel function of order 0 is just the popular $\hbox{sinc}$ function $\sin{x}/x$.} of order $n$ and $z_{ns}$ is the $s^{th}$ root of the $j'_n$ function.
The theoretical resonance frequencies are
\be
	f_{ns} = \frac{c}{2 \pi a} z_{ns}			\label{fns}
\ee
where $c$ is the propagation speed of sound and $a$ is the radius of the sphere \cite{Moldover86}.

The spherical Bessel functions $j_n$ can be studied by reference to the cylindrical Bessel functions $J_n$, thanks to the relation \cite{MorseAndIngard}
\be
	j_n(x) = \sqrt{\frac{\pi}{2x}} J_{n+\frac{1}{2}}(x) \; . 	\label{jn}
\ee 
The function $j'_n(x)$ has the following property:
\be
	j'_n(x) = \frac{1}{2n+1}[nj_{n-1}(x) - (n+1)j_{n+1}(x)]	 \; .	\label{djn}
\ee
By substituting (\ref{jn}) in (\ref{djn}) we get
\be
	j'_{n}(x) = \frac{1}{2n+1}\sqrt{\frac{\pi}{2x}}[nJ_{n-\frac{1}{2}}(x) - (n+1)J_{n+\frac{3}{2}}(x)] \; .
\label{derivate}
\ee

Some roots $z_{ns}$ of equation (\ref{equation}), i.e., the zeros of~(\ref{derivate}), are listed in Table~\ref{troots}. 
A closer look at the set of roots shows that they are not uniformly distributed, unlike the longitudinal resonances in tubes or between parallel boundaries. The roots are wider apart at low frequencies than at high frequencies. This effect is stronger for higher values of $n$ but, as it is clearly visible from Table~\ref{tPiroots}, any series of roots tend to be periodic in $\pi$ for high values of $s$. This can be interpreted as dispersion at low frequencies and will give us a hint on how to implement the spherical resonator.
\begin{table}[ht]
\begin{center}
	\begin{tabular}{|l|r|r|r|r|r|r|}
	\hline
	{\bf n $ \backslash $ s} & 1 &	2 &	3 &	4 &	5 & 	6	\\
	\hline
	0 &	0.00 &	4.49 &	7.73 &	10.90 & 14.07 & 17.22	\\
	1 &	2.08 &	5.94 &	9.21 &	12.40 & 15.58 & 18.74	\\
	2 &	0.00 &  3.34 &	7.29 &	10.61 &	13.85 & 17.04 	\\
	3 &	0.00 &  4.51 &	8.58 &	11.97 &	15.25 & 18.47 	\\
	4 &	0.00 &  5.65 &	9.84 &	13.30 &	16.61 & 19.86 	\\
	5 &	0.00 &  6.76 &	11.07 &	14.59 &	17.95 & 21.23   \\
	6 &	0.00 &  7.85 &	12.28 &	15.86 &	19.26 & 22.58   \\
	7 &	0.00 &  8.94 &	13.47 &	17.12 &	20.56 & 23.91   \\
	8 &	0.00 &  10.01 &	14.65 &	18.36 &	21.84 & 25.22 	\\
	9 &	0.00 &  11.08 &	15.82 &	19.58 &	23.11 & 26.52   \\
	\hline
	\end{tabular}
\end{center}
	\caption{Roots of $j'_n (x)=0$ for order $n=0,\dots,9$ 
			and root number $s = 1,\dots,6$.}
	 \label{troots}
\end{table}
\begin{table}[ht]
\begin{center}
	\begin{tabular}{|l|r|r|r|r|r|r|}
	\hline
	{\bf n $ \backslash $ s} &	2 &	3 &	4 &	5 & 	6	\\
	\hline
         0 &    1.4305 &    1.0287 &    1.0119 &    1.0065 &    1.0041 \\
         1 &    1.2283 &    1.0394 &    1.0181 &    1.0106 &    1.0069 \\
         2 &	1.0640 &    1.2566 &    1.0580 &    1.0289 &    1.0176 \\
         3 &    1.4370 &    1.2954 &    1.0787 &    1.0414 &    1.0261 \\
         4 &    1.7976 &    1.3349 &    1.0998 &    1.0548 &    1.0355 \\
         5 &    2.1508 &    1.3731 &    1.1206 &    1.0684 &    1.0453 \\
         6 &    2.4992 &    1.4096 &    1.1408 &    1.0821 &    1.0553 \\
         7 &    2.8442 &    1.4442 &    1.1604 &    1.0956 &    1.0654 \\
         8 &    3.1866 &    1.4772 &    1.1794 &    1.1089 &    1.0755 \\
         9 &    3.5268 &    1.5087 &    1.1977 &    1.1219 &    1.0854 \\
	\hline
	\end{tabular}
\end{center}
	\caption{Differences of contiguous roots of $j'_n (x)=0$ (normalized to $\pi$) for order $n=0,\dots,9$ 
			and root number $s = 2,\dots,6$.}
	 \label{tPiroots}
\end{table}
Table~\ref{tHzroots} shows the actual frequency values of the resonances of a sphere having radius $a = 0.188 \mbox{m}$, filled with air at the temperature\footnote{For the sake of accuracy, we take into account the fact that the propagation speed $c$ is temperature dependent according to the following relation:
\be
	c=331.8 \sqrt{\frac{t+273}{273}} \quad \quad (m/s)
\ee
where $t$ is the temperature $^\circ \mbox{C}$ of the air contained in the shape.} $t = 23^\circ \mbox{C}$. This particular enclosure will be used throughout the paper as a test case.
\begin{table}[ht]
\begin{center}
	\begin{tabular}{|l|r|r|r|r|r|r|}
	\hline
	{\bf n $ \backslash $ s} & 1 &	2 &	3 &	4 &	5 & 	6	\\
	\hline
         0 &  0.000 &   1.314 &   2.260 &   3.189 &   4.114 &   5.037 \\
         1 &  0.609 &   1.738 &   2.693 &   3.628 &   4.557 &   5.482 \\
         2 &  0.000 &   0.977 &   2.132 &   3.105 &   4.050 &   4.985 \\
         3 &  0.000 &   1.321 &   2.511 &   3.502 &   4.459 &   5.402 \\
         4 &  0.000 &   1.652 &   2.878 &   3.889 &   4.858 &   5.810 \\
         5 &  0.000 &   1.976 &   3.238 &   4.268 &   5.249 &   6.210 \\
         6 &  0.000 &   2.297 &   3.592 &   4.640 &   5.634 &   6.604 \\
         7 &  0.000 &   2.614 &   3.941 &   5.007 &   6.014 &   6.992 \\
         8 &  0.000 &   2.928 &   4.285 &   5.369 &   6.388 &   7.376 \\
         9 &  0.000 &   3.241 &   4.627 &   5.728 &   6.759 &   7.756 \\
	\hline
	\end{tabular}
\end{center}
	\caption{Resonance frequencies (in kHz) corresponding to roots of $j'_n (x) = 0$ for order $n=0,\dots,9$ and root number $s = 1,\dots,4$, for a sphere having radius $0.188\mbox{m}$ at a temperature of $23^\circ \mbox{C}$}
	 \label{tHzroots}
\end{table}
It should be noticed that the Bessel functions of any order give rise to a resonance at dc, except for order $n=1$. In the design procedure of section~\ref{design} we neglect this singularity and we treat all the Bessel orders as if they would provide a resonance at dc. This approximation produces inaudible effects, as humans are insensitive to very low frequencies.

Relation (\ref{fns}) yields the theoretical resonant frequencies of the sphere. Since the envelope of the sphere might be vibrating as well as dissipating some acoustic energy, these frequencies should be corrected for the effects that occur at the boundary~\cite{Moldover86,Kuttruff}.

\clearpage

\section{Measurements}
Data are available from 3 experiments: Moldover {\sl et al.}~\cite{Moldover86} display a spectrum measured in an argon-filled thick metal shell, and we have measured resonances in a rigid plastic shell as well as in an inflatable plastic ball.

In the experiment conducted on the metal shell, many roots $z_{ns}$ can be identified~\cite{Moldover86}. The series of roots $z_{1s}$ can be measured but the amplitudes are very small. The series $z_{0s}$ shows larger amplitudes but the largest amplitudes are found at $z_{22}$, $z_{42}$ and $z_{62}$. Due to this observation, these three important resonances were explicitely modelled by second-order filters in the early implementation of fig.~\ref{BasicSphere}~\cite{Dutilleux99}. In general, in the experiment done with the metal shell, the series $z_{2p,s}$ are well represented, whereas the series $z_{2p+1,s}$ have lower amplitudes. The relative difference in amplitude between the resonances is due to different sensitivity to wall absorption and to the position of transducers during the measurement.

\begin{figure}[ht]
\if T\draft
	\epsfxsize=6.0cm
\else
	\epsfxsize=3cm
\fi
	\centerline{\epsfbox{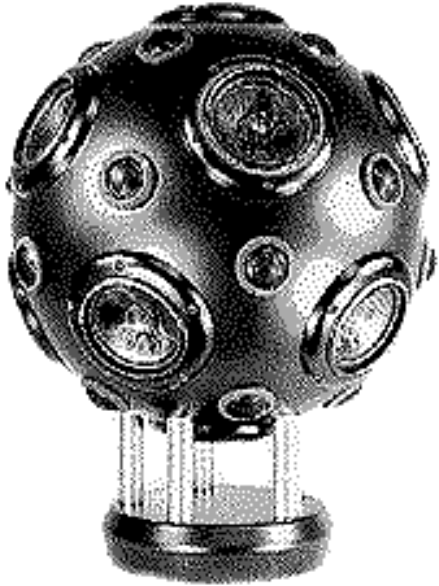}}
	\caption{The SAL Star-5 spherical loudspeaker.}
	\label{Star5}
\end{figure}

\begin{figure}[ht]
\if T\draft
	\epsfxsize=9.0cm
\else
	\epsfxsize=7.0cm
\fi
	\centerline{\epsfbox{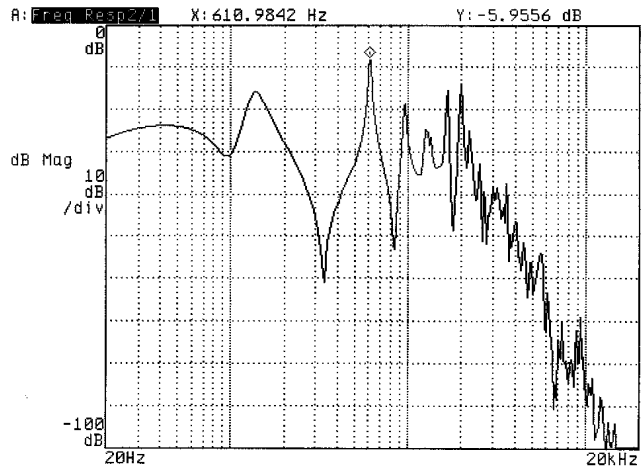}}
	\caption{Frequency response measured within the enclosure 
		of the spherical loudspeaker. $f_{11}$ can clearly be 
		identified at $611 Hz$.}
	\label{Star5-611}
\end{figure}

\begin{figure}[ht]
\if T\draft
	\epsfxsize=9.0cm
\else
	\epsfxsize=7.0cm
\fi
	\centerline{\epsfbox{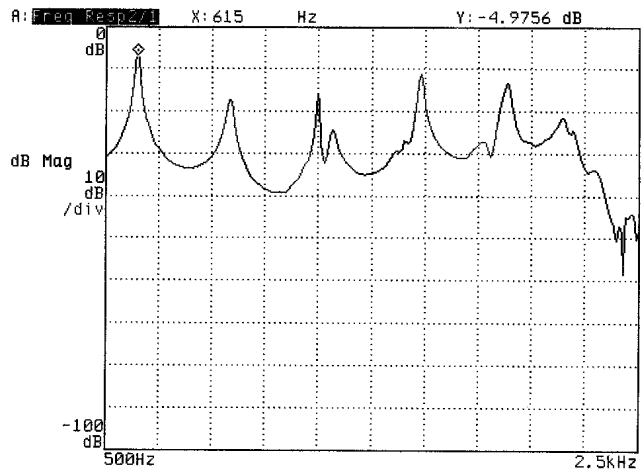}}
	\caption{Close up of fig.~\ref{Star5-611} between $500 Hz$
	 and $2500 Hz$. Six resonances can be identified.}
	\label{Star5-500-2500}
\end{figure}

The resonant frequencies of the spherical loudspeaker depicted in fig. 
\ref{Star5} were measured. In this loudspeaker 12 transducers are mounted on a spherical ABS plastic enclosure~\footnote{We thank Speaker Array Logic for providing the loudspeaker.}.
Several resonances could be identified with sufficient accuracy 
(see figures~\ref{Star5-611},~\ref{Star5-500-2500}
and Tab.~\ref{fStar5})~\cite{Schwoe}).

\begin{table}[ht] \label{fStar5}
	\begin{center}
	\begin{tabular}{|l|r|r|r|r|}
	\hline
	$f_{ns}$ & $f_m$ {\scriptsize (Hz)} & $f_{th}$ {\scriptsize (Hz)} &
	 	up {\scriptsize $\%$} \\
	\hline
	$f_{02}$ & 1290 & 1319 & -2	\\
	$f_{11}$ &  615 &  611 &  1	\\
	$f_{22}$ &  960 &  981 & -2	\\
	$f_{32}$ & 1350 & 1325 &  2	\\
	$f_{42}$ & 1680 & 1657 &  1	\\
	$f_{52}$ & 2000 & 1983 &  1	\\
	$f_{62}$ & 2240 & 2304 & -3	\\
	\hline
	\end{tabular}
	\caption{Resonances of the spherical loudspeaker Star-5. ($f_m$)
measured frequency, ($f_{th}$) is the theoretical value, (up) is the 
sharpness. The theoretical values are computed for a radius of $0.188 m$ and 
a temperature of $t = 25^\circ \mbox{C}$.}
	\end{center}
\end{table}

The resonances of an inflatable plastic ball having a diameter of $0.67 m$ were also measured by posing the plastic ball onto a small loudspeaker. The loudspeaker was playing test signals through the ball and a microphone recorded the sound filtered by the ball (fig.~\ref{realsetup}) at the temperature of $23^\circ \mbox{C}$ degrees. The position of the microphone was chosen in order to balance the amplitude of the various resonances. A spurious resonance showed up at $200 \mbox{Hz}$, probably due to the imperfect  coupling between the loudspeaker and the ball, so that it was necessary to prefilter the excitation signal with a parametric notch equalizer tuned at $200 \mbox{Hz}$.
	\Comment{A stethoscope in contact with the sphere is very useful to explore different locations but it fails at detecting the resonance $f_{11}$.} 

Fig.~\ref{realball} shows the frequency response measured with a white noise generator and a spectrum analyzer. We found that the low-frequency resonances are systematically sharper than the theoretical values (see table~\ref{fpball}). We assume that these deviations are  due to the compliance offered by the plastic boundary, which can not be considered as a rigid wall at low frequency. The prominent resonances could be identified with  $f_{11}$, $f_{22}$, $f_{32}$, $f_{42}$, $f_{52}$, 
$f_{62}$, $f_{72}$ and $f_{92}$.

\begin{table}[ht] 
	\begin{center}
	\begin{tabular}{|l|r|r|r|r|}
	\hline
	$f_{ns}$ & $f_m$ {\scriptsize (Hz)} & $f_{th}$ {\scriptsize (Hz)} & up {\scriptsize $\%$} \\
	\hline
	$f_{11}$ & 400 & 340 & 17.0	\\
	$f_{22}$ & 588 & 546 & 7.7	\\
	$f_{32}$ & 772 & 738 & 4.6	\\
	$f_{42}$ & 944 & 923 & 2.3	\\
	$f_{52}$ & 1120 & 1104 & 1.5	\\
	$f_{62}$ & 1306 & 1283 & 1.8	\\
	$f_{72}$ & 1470 & 1460 & 0.7	\\
	$f_{92}$ & 1810 & 1810 & 0.0	\\
	\hline
	\end{tabular}
	\caption{Resonances of the $0.67 m$ plastic ball. ($f_m$) is the measured frequency, ($f_{th}$) is the theoretical value, (up) is the sharpness. The theoretical values are computed for a diameter of $0.673 m$ and 
a temperature of $t = 23^\circ \mbox{C}$.
	\label{fpball}
	}
\end{center}
\end{table}

\clearpage

\section{Spherical resonator model}
\label{sphmodel}
The first attempt to imitate spherical resonators, as shown in 
figure~\ref{BasicSphere}, has provided audible results that allow users to distinguish 
simulated spheres from other shapes such as cube, quader or tube just from their 
specific sound colour. This seems to confirm experiments on object recognition performed with blind subjects~\cite{McGrath99}.
From the AML model, we have learnt that the fundamental 
frequencies ($f_{11}, f_{22}, f_{32}, \ldots , f_{n2}$) should be modelled with a 
high accuracy.
	The periodic spacing of resonances at higher frequencies 
is an effective parameter but it is difficult to tune. If all the 
resonances were spaced according to the asymptotic spacing $\pi$ of the $z_{ns}$ roots,	
they would imitate a cylinder rather than a sphere. On the other hand, the spacing at low 
frequencies is much larger than the asymptotic spacing (Tab.~\ref{tPiroots}), 
so the spacing was tuned empirically at an intermediate value.
	In the early experiment, the $z_{ns}$ were grouped according to the number $s$ of the roots 
	rather than to the order $n$ of the function because we noticed that the 
	spacing in the $z_{ns}$ series was more regular along $s$ than along $n$.

Building on the experience gathered with this first model, we try here to design a 
more systematic one.
Consider the perfectly-reflecting rectangular box and its representation in the BaBo model. Let us see how the model can be extended to spherical enclosures.

Indeed we would like to consider a perfectly reflecting sphere as a parallel connection of non-harmonic comb filters. The resonances of the $n^{th}$ comb filter correspond to the local extremal points of the $n^{th}$ order spherical Bessel function.

\begin{figure}[ht]
\if T\draft
	\epsfxsize=9.0cm
\else
	\epsfxsize=7.0cm
\fi
\centerline{\epsfbox{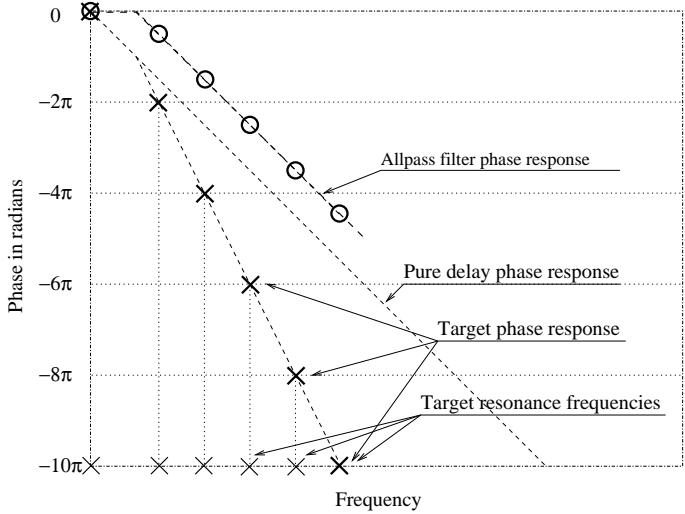}}
\caption{Simplified view of the phase response of the inharmonic comb filter loop associated with a Bessel function. It is shown how this phase response can be constructed as the sum of the contribution of a pure delay with the phase response of an allpass filter.}
\label{phaseCartoon}
\end{figure}

\begin{figure}[ht]
\if T\draft
	\epsfxsize=9.0cm
\else
	\epsfxsize=7.0cm
\fi
\centerline{\epsfbox{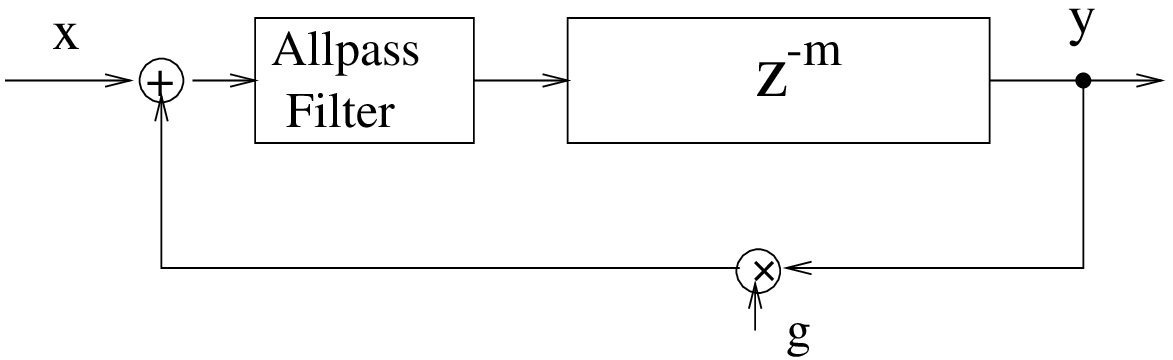}}
\caption{Inharmonic comb filter reproducing the modal resonances associated with a Bessel function of a given order}
\label{inhacomb}
\end{figure}

Given the resonance frequencies, we can sketch the ideal phase response of the $n^{th}$ comb filter loop, since the loop phase has to be equal to a multiple of $2\pi$ in order to sustain the mode associated with a resonance. Fig.~\ref{phaseCartoon} schematically shows with bold crosses the desired phase response at the positions of resonances, as it is generally found for any Bessel function order. A straight line fits the crosses fairly well for all Bessel orders, so that we can construct a realizable target phase response that is the sum of a linear contribution (realizable by a pure delay)  and a piecewise linear contribution (that can be approximated by a stable allpass filter). 
The scheme of one of the inharmonic comb filters is shown in fig.~\ref{inhacomb}, where the allpass filter and delay line are the main components of the feedback loop. In a practical implementation, one should also include a lowpass filter that accounts for frequency-dependent losses in a modal series. Such a filter can be designed starting from measured modal decay times. Alternatively, its parameters can be left open to user adjustment. In any case, we are neglecting losses at this stage, as we are mainly interested in the frequency distribution of resonances rather than in their relative strength.

\begin{figure}[ht]
\if T\draft
	\epsfxsize=9.0cm
\else
	\epsfxsize=7.0cm
\fi
\centerline{\epsfbox{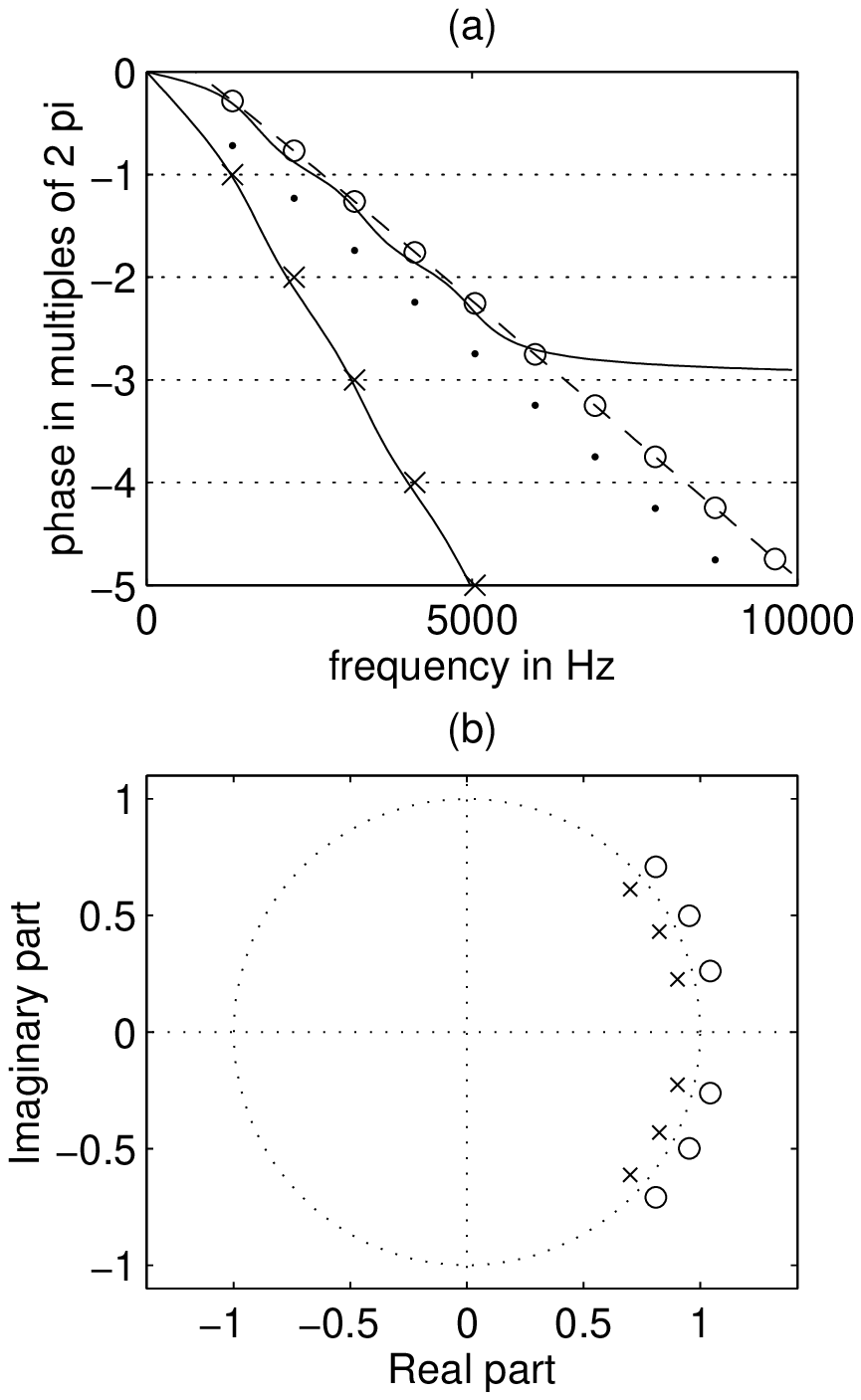}}
\caption{(a): Phase response of the feedback loop of a inharmonic comb filter reproducing the resonances of a spherical resonator ($r = 0.188 \hbox{m}$) associated with the Bessel function of order $0$: $\times$: phase response at resonance points; $\bullet$: phase provided by the delay ($8$ samples); $\circ$: target phase residue to be approximated by  the allpass filter; dashed line: polynomial curve approximating the target phase points; solid lines: designed allpass filter phase response and overall approximated phase response. (b): Pole-zero plot of the designed allpass filter.}
\label{phaseap}
\end{figure}

Indeed, in the allpass filter design stage, we do not need to approximate all the piecewise linear dashed line of fig.~\ref{phaseCartoon}. Instead, we just have to make sure that the designed phase response passes as close as possible to the points represented as small circles.
Fig.~\ref{phaseap}.a shows with crosses the phase response of the $0$ order feedback loop at the resonance points. A monotonic phase curve interpolating those points can be obtained as the sum of a linear ramp and a nonlinear residual, also shown in fig.~\ref{phaseap}.a with dots and circles, respectively. According to the scheme of fig.~\ref{inhacomb}, the linear component is given by a delay line whose length is equal to the slope of the linear ramp, and the nonlinear residual can be provided by an allpass filter.
 
A second design trade off appears here: up to which number $S$ should the 
$f_{ns,~0 \leq s \leq S}$ be approximated? Measurements and computations 
show that it is rather difficult to identify $f_{ns,~s > 1}$ resonances 
since slight inaccuracies suffice to change assignements. The experience gained 
in the first model (fig. \ref{BasicSphere}) indicates that $S$ can not be too small if the ``sense of roundness'' has to be maintained in the medium and high frequency range.
The order chosen for the allpass filter will determine the number $S$ of 
resonances that will be accurately imitated.

\begin{figure}[ht]
\if T\draft
	\epsfxsize=9.0cm
\else
	\epsfxsize=7.0cm
\fi
\centerline{\epsfbox{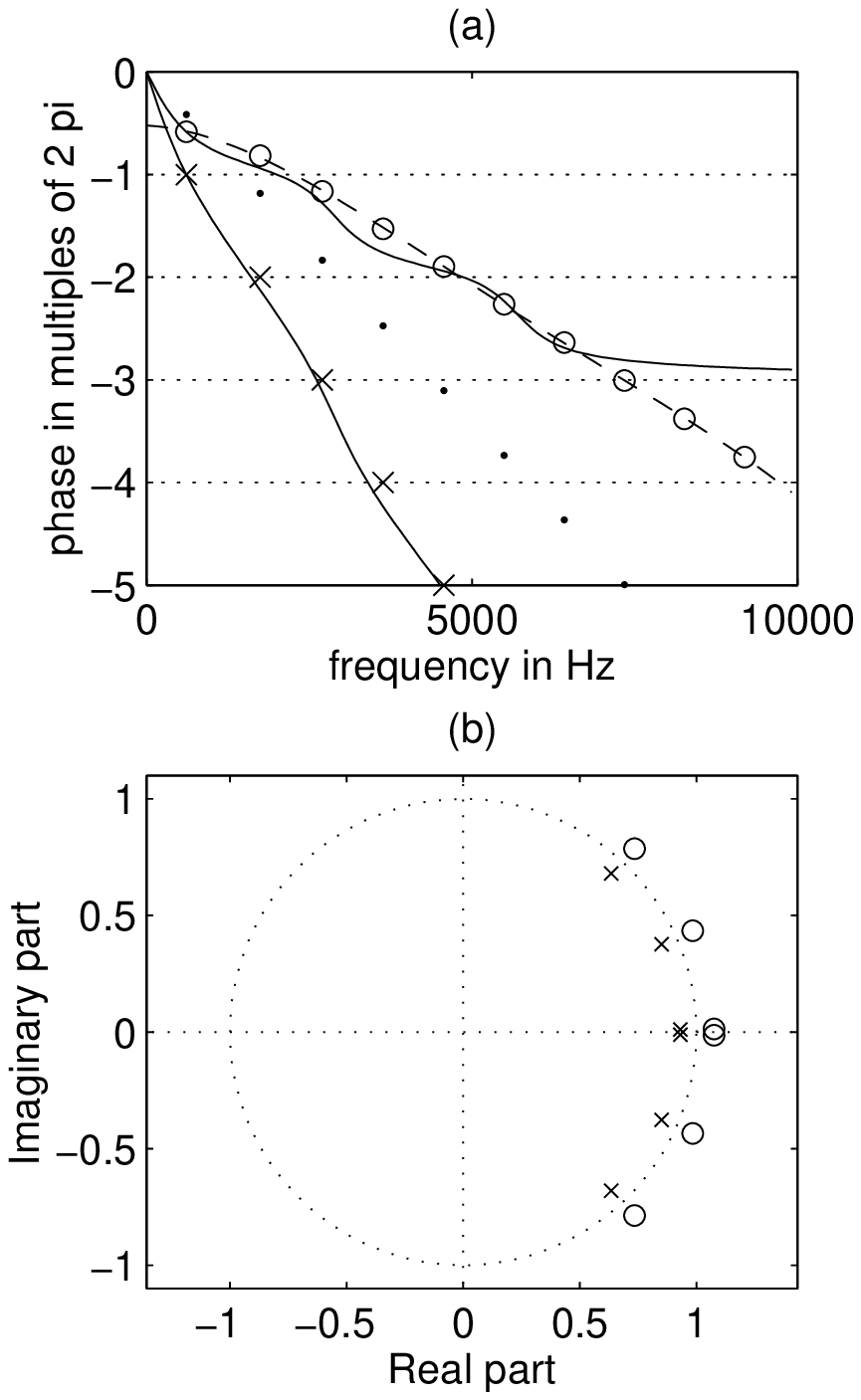}}
\caption{Same parameters as fig.~\ref{phaseap}. Bessel function of order 1.}
\label{phaseap1}
\end{figure}

\begin{figure}[ht]
\if T\draft
	\epsfxsize=9.0cm
\else
	\epsfxsize=7.0cm
\fi
\centerline{\epsfbox{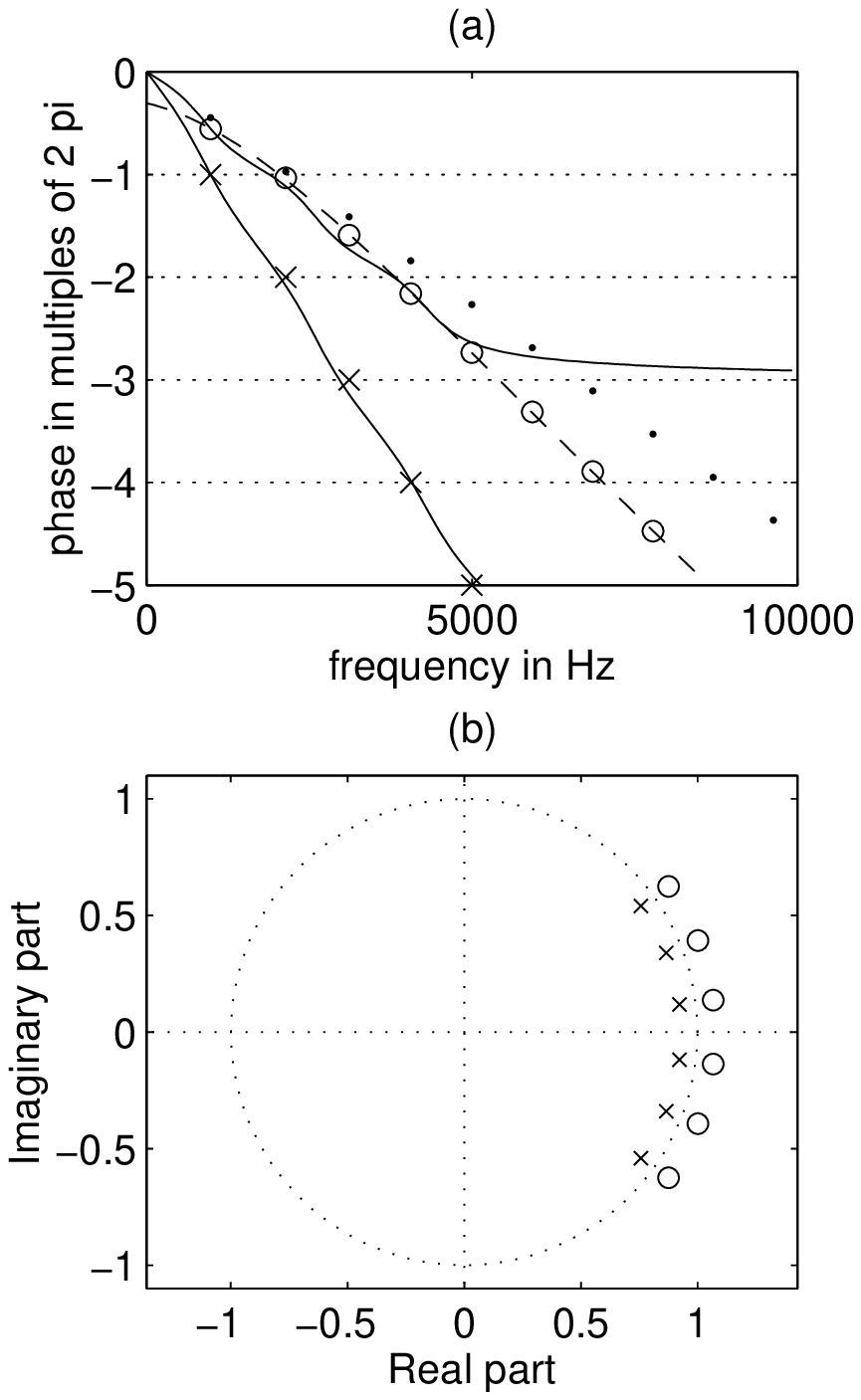}}
\caption{Same parameters as fig.~\ref{phaseap}. Bessel function of order 2.}
\label{phaseap2}
\end{figure}

\begin{figure}[ht]
\if T\draft
	\epsfxsize=9.0cm
\else
	\epsfxsize=7.0cm
\fi
\centerline{\epsfbox{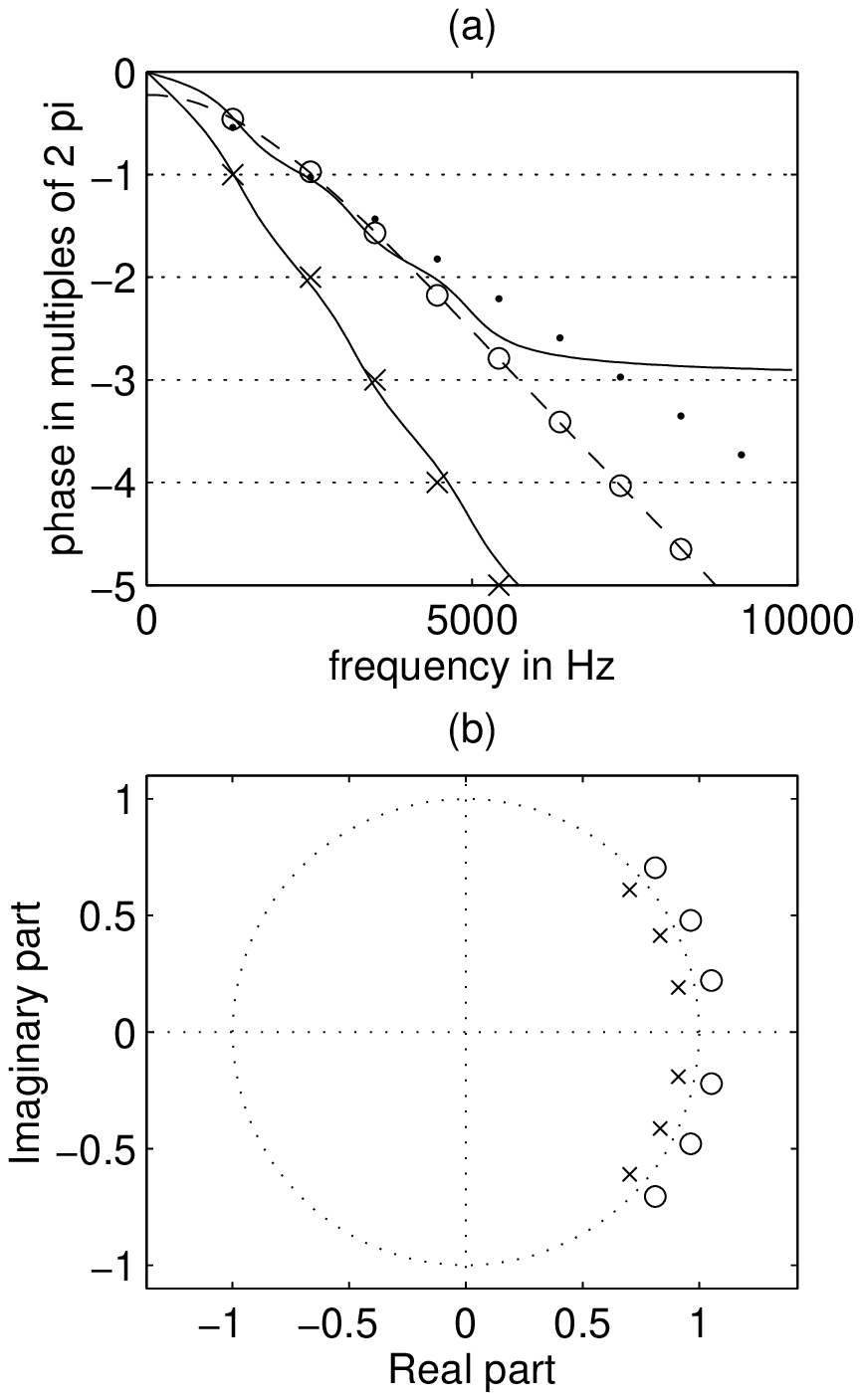}}
\caption{Same parameters as fig.~\ref{phaseap}. Bessel function of order 3.}
\label{phaseap3}
\end{figure}

\begin{figure}[ht]
\if T\draft
	\epsfxsize=9.0cm
\else
	\epsfxsize=7.0cm
\fi
\centerline{\epsfbox{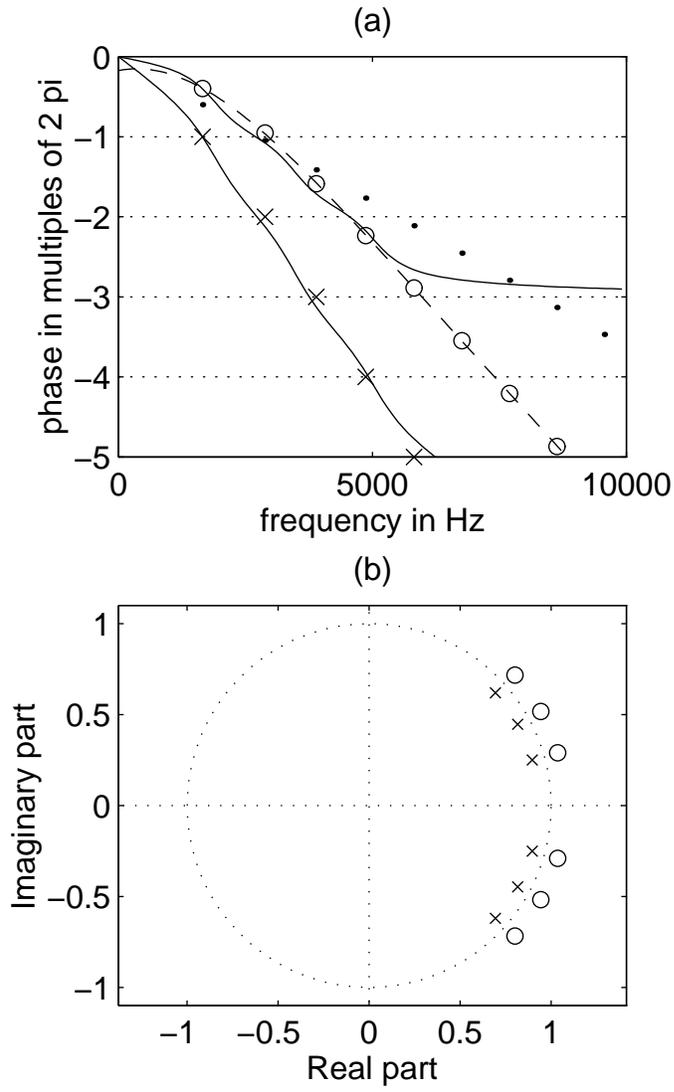}}
\caption{Same parameters as fig.~\ref{phaseap}. Bessel function of order 4.}
\label{phaseap4}
\end{figure}

The nonlinear phase curve can be roughly approximated by a couple of linear segments, a low-frequency slope and a high-frequency slope. With this observation, the allpass filter for the $0$ order Bessel function can be designed by placing the poles on the unit circle according to these two slopes. Fig.~\ref{phaseap}.b shows a zero-pole distribution that gives the two-slope phase response interpolating the small circles, which is depicted in solid line in fig.~\ref{phaseap}.a. 
Figures~\ref{phaseap1} to~\ref{phaseap4} show the phase responses and pole-zero plots for the same sphere as in fig.~\ref{phaseap}, but with a Bessel function order going from 1 to 4. Table~\ref{besstab} shows the designed parameters for radius equal to $0.188 \mbox{m}$ and $0.32 \mbox{m}$,  and Bessel functions going from 0 to 2. Reasonably shaped allpass phase responses can be obtained just by controlling two parameters, the argument of the first (low-frequency) pole, and the angular distance between any couple of contiguous poles.

\clearpage

\section{Design Procedure and Examples}
\label{design}
We have written a design procedure 
\Comment{\footnote {The Matlab script {\tt iterar.m}}} 
that finds the combinations of allpass filter and delay length for a set of sphere radii by iterative optimization over the position of the knee and the second, largest slope of the phase curve. In other words, the procedure  iteratively changes the angle of the low-frequency pole and the relative distance between all the other equidistant poles. The relative contribution of the delay line to the loop phase response is also subject to optimization. The distance of the poles from the centre is kept fixed as it is mainly responsible for the magnitude of the ripples in the phase response. This pole radius plays a minor role because we are not interested in the phase values between the positions of resonance peaks and, therefore, we are not interested in a particular value of the phase ripples.
A criterion for choosing the initial inter-pole distance is that the asymptotic distance between the peaks associated with a Bessel function is equal to a period of the ripple. The error that the iterative design procedure tries to minimize is the sum of the square phase differences at the target resonance points, weighted by some monotonically decreasing function that allows a more accurate positioning of resonances in low frequency.

If the goal is having a spherical resonator model that can be controlled in its radius, the design procedure should be run several times, each for a selected value of radius. In this way, a table can be constructed with a complete set of parameters for the resulting inharmonic comb filters. Notice that, even though a sixth-order allpass filter has six coefficients, the observation made in section~\ref{sphmodel} allows us to implement it as three second-order sections that can be controlled by two parameters: the angle of the first pole, and the angular distance of the following poles. The results of the design procedure for two distinct radii are reported in table~\ref{besstab}.
Since the precise position of resonances is of some importance only in the first few thousands Hz (say, $4 \hbox{kHz}$)~\cite{rocscaieee}, we see that a low order allpass filter is adequate for small spheres, e.g. order $6$ or less for radii smaller than $0.5 \hbox{m}$. 
However, for larger spheres, the filter order should be increased in order to provide a decent approximation at least under the first kHz. 

\Comment{
\begin{table*}
\begin{center}
\begin{tabular}{|l|l|l|l|l|l|l|}
\hline
radius [m] & 0.2  & 0.3  & 0.4 &  0.5 & 0.6 & 0.7 \\
knee [rad/s] & 0.54 & 0.36 & 0.27 & 0.22 & 0.18 & 0.16 \\ 
slope ($\times 10^{-3}$) & -1.7 & -2.6 & -3.5 & -4.4 & -5.3 & -6.2 \\
angle $1^{\hbox{st}}$ pole & 0.56  & 0.38 & 0.28 & 0.23 & 0.19 & 0.16 \\
separ. angle & 0.51 & 0.32 & 0.23 & 0.18 & 0.14 & 0.11 \\
delay [samples] & 4 & 6 & 8 & 10 & 12 & 14 \\
\hline
\end{tabular}
\caption{\sl{Parameters for the inharmonic comb filter of the 0 order Bessel function.}}
\end{center}
\label{bess0tab}
\end{table*}
}

\begin{table*}
\begin{center}
\begin{tabular}{|l|l|l|l|l|l|l|}
\hline
radius & knee & slope & angle & separ. & delay & fund. freq. \\
$[m]$ & $[rad/s]$ & ($\times 10^{-16}$) & $1^{\hbox{st}}$ pole & angle &  &  $[kHz]$\\
\hline 
\multicolumn{7}{l}{Bessel order 0}\\
\hline
{\bf 0.188} & 0.19  & -0.04 & 0.24 & 0.23 & 23 & 1.32\\
{\bf 0.32} & 0.11 & -0.05 & 0.20 & 0.15 & 48 & 0.77 \\
\hline 
\multicolumn{7}{l}{Bessel order 1} \\
\hline
{\bf 0.188} & 0.09 & -0.33 & 0.01 & 0.40 & 28 & 0.61 \\
{\bf 0.32} & 0.05 & -0.32 & 0.01 & 0.22 & 49 & 0.36 \\
\hline
\multicolumn{7}{l}{Bessel order 2} \\
\hline
{\bf 0.188} & 0.14 & -0.38 & 0.12 & 0.24 & 19 & 0.98\\
{\bf 0.32} & 0.08 & -0.34 & 0.16 & 0.25 & 53 & 0.58 \\
\hline
\end{tabular}
\caption{Parameters for the inharmonic comb filter. Bessel functions of order 0 to 2.}
\end{center}
\label{besstab}
\end{table*}

\begin{figure}[ht]
\if T\draft
	\epsfxsize=11.0cm
\else
	\epsfxsize=9.0cm
\fi
\centerline{\epsfbox{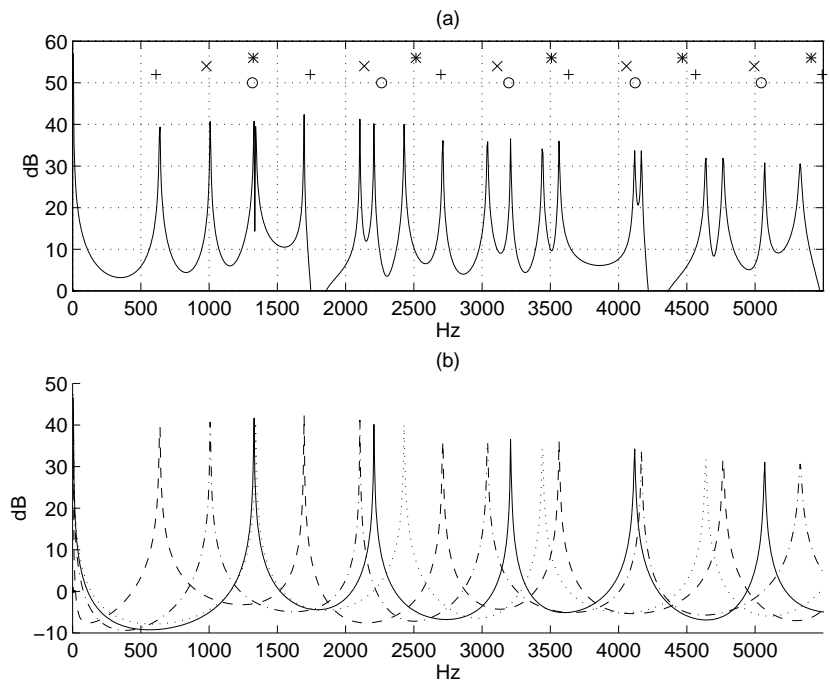}}
\caption{(a): Frequency response of the parallel of dispersive comb filters designed for radius 0.188. Resonance positions of the ideal sphere are indicated for Bessel functions of order 0 (o), 1 (+), 2 ($\times$), and 3 (*). (b): Superposition of the responses  of dispersive comb filters for Bessel functions of order 0 (solid), 1 (dashed), 2 (dash-dotted), and 3 (dotted). }
\label{freqresp}
\end{figure}

Fig.~\ref{freqresp}.a shows the frequency response of the parallel connection of dispersive comb filters, here designed for radius $0.188 \mbox{m}$. The crosses represent the ideal modal positions for Bessel functions of order 0 to 4. In order to have a good match between the first resonance of each comb and its theoretical position, we properly shaped a weighting function to be used in the iterative optimization procedure. Maximum weight is used around the first resonance, while the following resonances become gradually less important. Psychoacoustic investigations should be conducted in order to better understand if the approximations introduced can be perceived and if they affect the perceived object shape. However, informal listening seems to indicate that significant deviations from the theoretical partial positions can be tolerated without loosing the ``sense of roundness''. For instance, the sharpness of resonances measured in the plastic ball, and reported in Table~\ref{fpball}, does not seem to prevent the listener from identifying the enclosure as a sphere.

Fig.~\ref{freqresp}.b shows how the comb filters from order 0 to 3 separately contribute to the response of fig.~\ref{freqresp}.a, which is obtained by pure summation of the comb outputs. Around some resonances different modes coming from different Bessel series interact with each other, and the local result is either a magnification or an attenuation of the peak.
As well as with actual enclosures where the shape of the frequency response is dependent on the positions of exciter and pickup, with the FDN we can vary the shape of the response, without moving the resonances, just by changing the input and output coefficients~\cite{cmj95}, indicated as $b_i$ and $c_i$ in fig.~\ref{fdn}.

\section{Trials, Discussion, and Tuning}
The ability of a model to simulate actual objects can be checked by 
comparing sounds played through the real objects with sounds processed by 
the model.
	Let us consider two different objects: a cylinder and a ball. Their shapes are 
	far enough from each other to allow for a clear difference between processed sounds.

\begin{figure*}[ht]
\if T\draft
	\epsfxsize=11.0cm
\else
	\epsfxsize=9.0cm
\fi
\centerline{\epsfbox{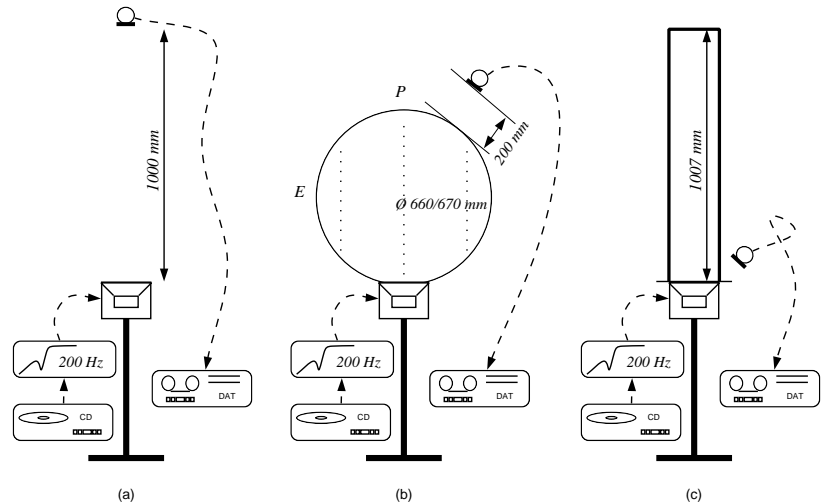}}
\caption{Measurement setup used for fig.~\ref{realball}. 
	A sound is played from a CD-player through a loudspeaker. On top of the 
	louspeaker, a sphere (b) or a tube (c) are posed. The sound modified by 
	the shape, acting as a resonator, is recorded by a microphone. 
	As a reference, the sound without any resonator is also recorded (a).
	A notch filter tuned at $200 Hz$ removes the disturbing resonance that appears when 
	the sphere is coupled to the loudspeaker.
	}
\label{realsetup}
\end{figure*}

Figure~\ref{realsetup} shows the setup used to record sounds processed 
by actual objects. The plastic ball is layed onto a small loudspeaker fed 
by a CD-player. Since the coupling between loudspeaker and sphere provoques 
a strong resonance at $200 Hz$, a notch filter is inserted between the 
sound source and the loudspeaker. The microphone picks up the sound 
radiated in the recording room. In a similar fashion, sounds played in 
a tube closed at one end are recorded. Since the transducers and the room 
are not completely neutral, a third recording is made that serves as a 
reference: the sound played through the loudspeaker alone. 

	In order to compare the sounds 
played through the sphere to those played through the tube, some kind of 
tuning is necessary. Tuning a tube is easy but tuning a sphere is definitely 
not, because its resonances are not harmonic from each other. 
Several tuning methods have been considered. 
One could tune the lowest resonance of the sphere 
to that of the tube but that would lead to sounds located in different registers. 
We have chosen a tuning method that adjusts the brightness of the sounds.
 Another possible method would 
be to ajust the volume of the objects. This last method is not suitable here
because a long tube with a small diameter will sound dramatically 
different from a short tube with a large diameter but having the same 
volume. This tuning is however meaningful for the comparison of a cube and a 
sphere. The brightness of the sounds played through the tube has been adjusted by 
modifying the length of the tube until it empirically fits the brightness 
of the sounds played through the sphere.

Some listeners  easily recognise the shape of the object just by listening 
to their sounds~\cite{Dutilleux99}. Most listeners agree that each object has a very peculiar 
sound colour but they fail at describing the differences and at 
telling one shape from the other. Many of them are capable of making a 
decision after a few learning trials, as if the knowledge had been available in 
their brain for a long time (perhaps since childhood) but that this 
knowledge had to be reactivated.

Similar tests were performed with sounds proces\-sed through the computer 
models. The FDN model of the sphere produces sounds with a refined reverberation, but the typical character of 
the sphere is less prominent. When comparing with FDN simulations of cubes, 
it is more difficult to tell the shape that is currently being simulated even though prior exposure still helps significantly. The fact that a sophisticated model performs worse than a simpler model for specific cases must be interpreted thinking that the initial model of the  sphere (fig.~\ref{BasicSphere}) was constructed for a specific radius by fine tuning of a few prominent resonances. On the other hand, the FDN model aims at achieving generality and complex behaviour with a compact parametric structure. However, further research in sound perception has to be pursued in order to understand which are the salient characteristics of a sound spectrum that bring us its shape signature. When such results become available, the design procedure of the FDN spherical model will be improved further.

\subsection{Including deviations from the ideal, rigid sphere}
\begin{figure}[h]
\if T\draft
	\epsfxsize=11.0cm
\else
	\epsfxsize=9.0cm
\fi
\centerline{\epsfbox{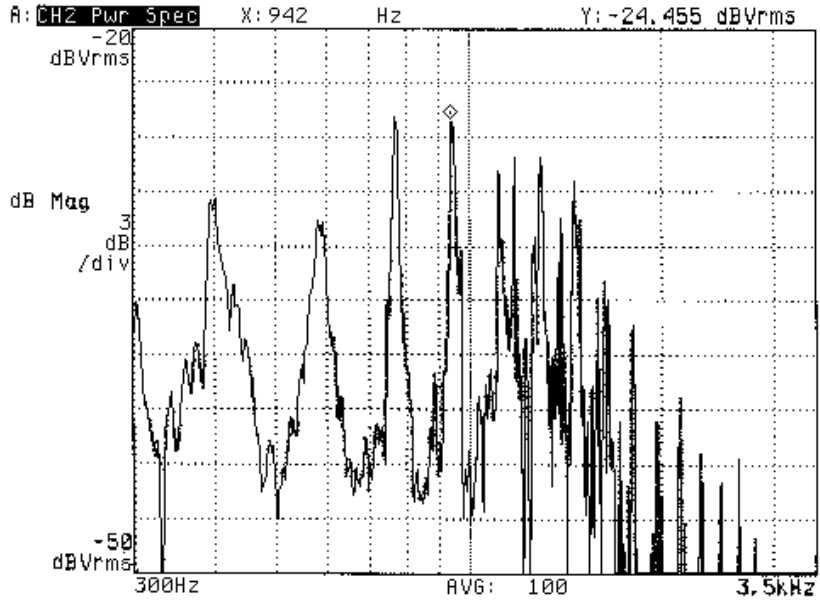}}
\caption{Measured frequency response of the plastic ball\Comment{ of fig.~\ref{realsetup}}.}
\label{realball}
\end{figure}

\begin{figure}[h]
\if T\draft
	\epsfxsize=12cm
\else
	\epsfxsize=9.0cm
\fi
\centerline{~~~~~~~\epsfbox{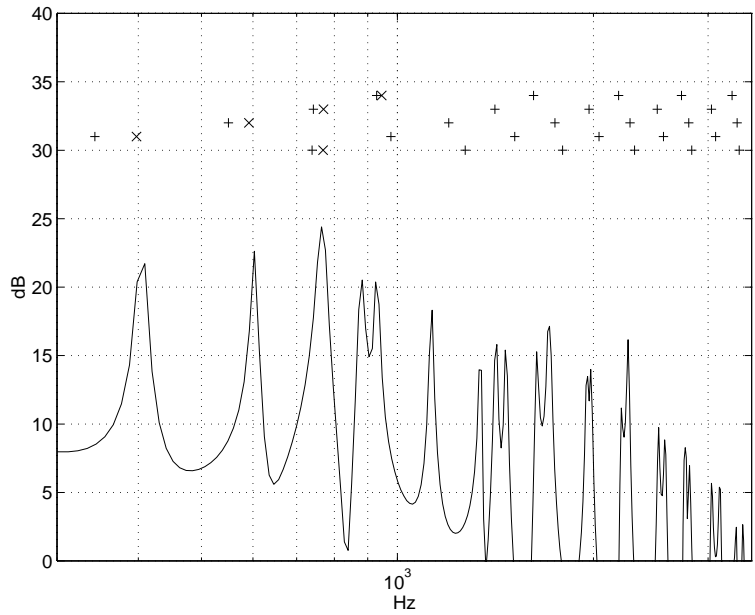}}
\caption{Frequency response of the FDN model of the plastic ball\Comment{ of fig.~\ref{realsetup}}. $+$: resonance positions of the ideal sphere. $\times$: measured resonance positions.}
\label{frtuned}
\end{figure}

The FDN has been optimized according to the theory of the sphere but, in order to compare it with the plastic ball, deviations from the theoretical tuning had to be implemented. Such deviations (see table~\ref{fpball}) can be introduced in our model just by moving the theoretical resonance positions in the procedure for designing the allpass filters. This can be done fairly easily if the deviations are small, otherwise it can be difficult to assign a certain resonance to an inharmonic comb series. Alternatively, one can start with the filters designed for the ideal sphere and adjust the position of the first pole, as we did to obtain the frequency response of fig.~\ref{frtuned}, which should be compared with fig.~\ref{realball}. In the feedback loop, we used second-order FIR filters (exhibiting a one-sample delay) to simulate the faster attenuation of higher modes. Moreover, a first-order lowpass filter has been cascaded with the whole structure in order to resemble the lowpass characteristic of fig.~\ref{realball}.

This tuning of the model 
improves the simulation but it is not yet enough to allow for a reliable 
detection of the simulated shape from the sound itself. Again, it seems that 
the physical theory of the sphere is not sufficient to drive the design stage,  and perceptual issues should be considered.
To this end, the issue of identifying a shape from its audible signature should 
be investigated more deeply in the future.

\Comment{In order to test the ability of the model to simulate actual objects, we have processed sounds through the FDN and have listened to the output and compared it with sounds that were recorded through the plastic ball (fig.~\ref{realsetup}). }

\clearpage

\section{Conclusion and Further Research}

We have shown that the basic structure of  the BaBo model can also be used to simulate  spherical geometries, just by using properly designed allpass filters within the filtering blocks $H_i$ of fig.~\ref{fdn}. The resulting model might be called ``Ball within the Ball'' (BaBa). We have shown that a simple design procedure and, possibly, some manual tweaking, allow an efficient structure to be realized that can be tuned either like an ideal sphere or like a real one. An open question is nevertheless how accurate this design has to be, since it is still unsettled which are the relevant parameters that affect the ``perceived roundness'' of the object. 

An important aspect of the model is that very few parameters are added to the BaBo model to control the allpass filters of the spherical model. Even deviations from ideal boundary conditions are reasonably achieved by moving only the position of one pole per inharmonic series. So far, we have simulated inharmonic series given by Bessel functions of order ranging from 0 to 6. In many cases, higher order inharmonic series are needed to achieve realism, but the fundamental resonance of those series seems to be most important while the higher resonances get drowned in the dense mixture of resonances from other series and their position is out of the bandwidth of perceived inharmonicity. So, we suggest to implement higher-order series as harmonic comb filters tuned to the fundamental frequency of that series. Some form of delay interpolation~\cite{LaaksoB} might be needed to position such fundamental frequencies with sufficient accuracy.

The same structure used for spherical resonators might be as well used for cylindrical resonators. In this case there is a harmonic series given by longitudinal modes, superimposed with inharmonic series of resonances whose positions are determined by the extremal points of cylindrical Bessel functions. Therefore, the feedback delay network should have the first delay line accounting for the longitudinal modes, and the remaining lines, cascaded with properly-designed allpass filters, accounting for the inharmonic, transversal modal series.

If the inharmonic series, each corresponding to a Bessel function of a certain order, are recreated by comb filters having a delay line and an allpass filter in the feedback loop, it is conceivable to control the degree of ``roundness'' of the enclosure by changing the relative contribution to the overall phase response given by the delay and by the allpass filter. Namely, if all the delays are increased we gradually move from a sphere to a cube with rounded faces. This continuous shape control, as well as the extension of the BaBo model to cylindrical shapes will be covered in future research. 


\newpage

\listoffigures

\newpage

\listoftables

\end{document}